\documentclass[english,aps, twocolumn, superscriptaddress, longbibliography]{revtex4-1}
\usepackage[T1]{fontenc}
\usepackage[latin9]{inputenc}
\usepackage{textcomp}
\usepackage{amsmath}
\usepackage{amssymb}
\usepackage{graphicx}
\usepackage{esint}

\makeatletter

\providecommand{\tabularnewline}{\\}

\@ifundefined{textcolor}{}
{%
 \definecolor{BLACK}{gray}{0}
 \definecolor{WHITE}{gray}{1}
 \definecolor{RED}{rgb}{1,0,0}
 \definecolor{GREEN}{rgb}{0,1,0}
 \definecolor{BLUE}{rgb}{0,0,1}
 \definecolor{CYAN}{cmyk}{1,0,0,0}
 \definecolor{MAGENTA}{cmyk}{0,1,0,0}
 \definecolor{YELLOW}{cmyk}{0,0,1,0}
}

\makeatother

\usepackage{babel}
\begin{document}

\title{Coupled Mode Theory for Semiconductor Nanowires}

\author{Robert Buschlinger}

\affiliation{{\small{}\textsuperscript{}Institute of Condensed Matter Theory and Optics, Abbe Center of Photonics, Friedrich Schiller University of Jena, Max-Wien-Platz 1, 07743 Jena, Germany}}

\author{Michael Lorke}

\affiliation{{\small{}\textsuperscript{}Institute for Theoretical Physics, University of Bremen, 28359 Bremen, Germany}}

\author{Ulf Peschel}

\affiliation{{\small{}\textsuperscript{}Institute of Condensed Matter Theory and Optics, Abbe Center of Photonics, Friedrich Schiller University of Jena, Max-Wien-Platz 1, 07743 Jena, Germany}}

\begin{abstract}
We present a model to describe the spatiotemporal evolution of guided modes
in semiconductor nanowires based on a coupled mode formalism. Light-matter interaction is modelled based on semiconductor
Bloch equations, including many-particle effects in the screened Hartree-Fock approximation. Appropriate boundary conditions are used to incorporate reflections at waveguide endfacets, thus allowing for the simulation of nanowire lasing. We compute the emission characteristics and temporal dynamics of $CdS$ and $ZnO$ nanowire lasers and compare our results both to Finite-Difference Time-Domain simulations and to experimental data. Finally, we explore the dependence of the lasing emission on the nanowire cavity and on the materials relaxation time.

\end{abstract}
\maketitle

\section{Introduction}

An increasing demand for fast communication technologies and the limitations
inherent to electronic integrated circuits has stimulated the
research on nanophotonic components. In particular semiconductor nanowires
have gathered widespread interest due to their simple fabrication
and their remarkable photonic properties, which allow them to act
as efficient waveguides and as resonators either for photonic and
plasmonic lasing or for harnessing polaritonic effects.\cite{duansingle2003, oultonplasmon2009, Sidiropoulos2014,  saxenaoptically2013, doi:10.1021/nl401355b} 

Semiconductor nanowires are complex photonic systems supporting multiple
longitudinal as well as transverse modes interacting through the nonlinear
response of the medium. The optical properties of the medium are influenced
by many-body effects of the excited carriers, which give rise to excitonic
absorption peaks and the appearance of polaritons in the weakly excited
regime. In the strongly excited regime, effects like phase-space filling,
screening and excitation-induced dephasing dominate the optical response,\cite{ElSayed:94b,Manzke:02,chow1999semiconductor,haug2004quantum}
giving rise to a broad gain profile. In order to make correct predictions
across the different regimes of excitation conditions, theoretical
models of light-matter interaction in semiconductor nanowires need
to incorporate all these effects.

Recently, we proposed a coupled Finite-Difference Time-Domain (FDTD)\cite{Yee66numericalsolution,taflove:2005}
and semiconductor Bloch equations (SBEs)\cite{haug2004quantum,chow1999semiconductor}
approach to the modelling of light-matter interaction in arbitrary semiconductor geometries\cite{PhysRevB.91.045203}.
A similiar approach has also been applied to the description of semiconductor quantum wells.\cite{PhysRevB.94.115303}
While models based on the FDTD method are the most general, the numerical complexity
can be decreased considerably if assumptions are made concerning the
simulated geometry. 

In the present case we deal with nanowires, where
the propagating fields can be decomposed into waveguide eigenmodes. 
A considerable simplification of the numerical treatment is achieved by describing  
the evolution of the eigenmode amplitudes in the framework of coupled mode theory (CMT). In the general case a transverse
resolution is needed, but considerably less data points than in the case of a more general approach like FDTD are necessary. 
The resulting reduction in computational demands makes it possible to increase the simulated time window and the size and complexity of the nanowire geometry or to include even more sophisticated material models capturing additional effects relevant for semiconductor lasers as comprehensively summarized in \cite{Boehringer2008159,Boehringer2008247}.

In this paper, we describe our coupled mode theory for semiconductor
nanowires including the treatment of reflecting endfacets. We apply
the model to the simulation of the temporal dynamics of $ZnO$ nanowire-lasers
and compare our results to data from an experimental study\cite{0957-4484-27-22-225702}.

\section{Theoretical model}

\subsection{Derivation of propagation equations}

In the following, we summarize the derivation of evolution equations
for the slowly varying envelopes of waveguide modes under the influence
of material nonlinearities and dispersion, which have been used in
similiar form by other authors.\cite{Crosignani:81,snyder1983optical}
We start with Maxwell's equations in the frequency domain
\begin{equation}
\nabla\times\vec{E}=i\omega\mu_{0}\vec{H}\label{eq:maxwell1-1}
\end{equation}
\begin{equation}
\nabla\times\vec{H}=-i\omega\varepsilon_{0}\varepsilon\vec{E}-i\omega\vec{P}.\label{eq:maxwell2-1}
\end{equation}
The polarization term $\vec{P}$ couples the material
model including dispersion, absorption and nonlinearity to the equations for the
electromagnetic fields. We consider a waveguide extending in the $z$-direction,
which in the unperturbed case ($\vec{P}=0$) supports modes with the
transverse field profiles $\vec{E}_{m}^{\pm}\left(x,y\right)$, $\vec{H}_{m}^{\pm}\left(x,y\right)$
and the propagation constants $\pm\beta_{m}$. Neglecting group velocity
dispersion, the propagation constant close to the frequency $\omega_{0}$
takes the form 
\begin{equation}
\beta_{m}=\beta_{0,m}+\frac{1}{v_{m}}\left(\omega-\omega_{0}\right),\label{eq:BetaDisp}
\end{equation}
with $\beta_{0,m}=\beta_{m}\left(\omega_{0}\right)$ and $\frac{1}{v_{m}}=\left[\frac{\partial\beta_{m}}{\partial\omega}\right]_{\omega_{0}}$
being the modes inverse group velocity at $\omega_{0}$. Note, that by
using this simplification we merely neglegt the group velocity dispersion
arising from the frequency-dependent changes of the mode shape. Any dispersion
effects caused by the resonant excitation of the semiconductor material remain unaffected.

The fields propagating in the perturbed waveguide ($\vec{P}\neq0$)
can be decomposed into propagating modes as
\begin{equation}
\vec{E}=\sum_{\pm,m}u_{m}^{\pm}(z)\vec{E}_{m}^{\pm}\label{eq:modes1}
\end{equation}
\begin{equation}
\vec{H}=\sum_{\pm,m}u_{m}^{\pm}(z)\vec{H}_{m}^{\pm}.\label{eq:modes2}
\end{equation}
 We now analyze the vector $\vec{\Gamma}_{m}=\vec{E}\times\vec{H}_{m}^{\pm*}e^{\mp i\beta_{m}z}+\vec{E}_{m}^{\pm*}e^{\mp i\beta_{m}z}\times\vec{H}$,
which can be interpreted as the contribution of mode $m$ to the Poynting
vector. Using equations \eqref{eq:maxwell1-1} and \eqref{eq:maxwell2-1},
we find that 
\begin{equation}
\operatorname{div}\vec{\Gamma}_{m}=i\omega\vec{E}_{m}^{\pm*}\vec{P}e^{\mp i\beta_{m}z}.\label{eq:Gamma}
\end{equation}
Next we integrate equation $\eqref{eq:Gamma}$ with respect to the transverse
coordinates $x$ and $y$. As $\vec{\Gamma}_{m}$ decays exponentially with $x$ and $y$ approaching infinity, we obtain 
\begin{multline}
\int\limits_{-\infty}^\infty\int\limits_{-\infty}^\infty\frac{d}{dz}\left(\left[\vec{E}\times\vec{H}_{m}^{\pm*}+\vec{E}_{m}^{\pm*}\times\vec{H}\right]_{z}e^{\mp i\beta_{m}z}\right)dxdy\\
=i\omega\int\limits_{-\infty}^\infty\int\limits_{-\infty}^\infty\left(\vec{E}_{m}^{\pm*}\vec{P}e^{\mp i\beta_{m}z}\right)dxdy\label{eq:Intermediate}
\end{multline}
Next, we insert the mode expansion \eqref{eq:modes1},\eqref{eq:modes2} into equation
\eqref{eq:Intermediate}. Due to the orthogonality of the modes, we
obtain the expression 
\begin{equation}
\frac{d}{dz}u_{m}^{\pm}\mp i\beta_{m}u_{m}^{\pm}=\frac{i\omega}{4\gamma_{m}}
\int\limits_{-\infty}^\infty\int\limits_{-\infty}^\infty\left(\vec{E}_{m}^{\pm*}\vec{P}\right)dxdy\label{eq:evolutiondz}
\end{equation}
using the guided power of mode $m$ defined as
\begin{equation}
\gamma_{m}=\frac{1}{4}
\int\limits_{-\infty}^\infty\int\limits_{-\infty}^\infty\left(\vec{E}_{m}^{\pm}\times\vec{H}_{m}^{\pm*}+\vec{E}_{m}^{\pm*}\times\vec{H_{m}}\right)_{z}dxdy.
\end{equation}
We insert equation \eqref{eq:BetaDisp} and transform equation \eqref{eq:evolutiondz}
back to the time-domain using the inverse Fourier transforms $\vec{P}(z,t)=\mathcal{F}^{-1}\left\{ \vec{P}\right\} $ and
$u\left(z,t\right)=\mathcal{F}^{-1}\left\{ u\right\} $. We further define slowly varying envelopes
for the mode amplitudes
\begin{equation}
\hat{u}\left(z,t\right)=u\left(z,t\right)e^{i\omega_{0}t-i\beta_{0}z}
\end{equation}
 and for the polarization 

\begin{equation}
\hat{\vec{P}}(z,t)=\vec{P}(z,t)e^{i\omega_{0}t},
\end{equation}
and arrive at the evolution equation 
\begin{equation}
\pm\frac{1}{v_{m}}\frac{\partial}{\partial t}\left(\hat{u}_{m}^{\pm}\right)=-\frac{d}{dz}\hat{u}_{m}^{\pm}+De^{\mp i\beta_{0m}z}\label{eq:modeprop}
\end{equation}
for the slowly varying envelopes with the driving term 
\begin{equation}
D=-\frac{\frac{\partial}{\partial t}-i\omega_{0}}{4\gamma_{m}}
\int\limits_{-\infty}^\infty\int\limits_{-\infty}^\infty\left(\vec{E}_{m}^{\pm*}\hat{\vec{P}}\right)dxdy.\label{eq:driver}
\end{equation}
The self-consistent electric field driving the polarization used in
equation \eqref{eq:driver} is given as a superposition of all modes, taking into account the individual mode shapes. For
some applications it can be desirable to include a pump field $\vec{E}_{pump}\left(\vec{r},t\right)$,
which mainly propagates in directions perpendicular to the waveguide
axis and therefore does not contribute to the guided modes directly.
Thus, the self-consistent field is described as 
\begin{gather}
\vec{E}(\vec{r},t)=\sum_{\pm,m}\hat{u}_{m}^{\pm}(z,t)\vec{E}_{m}^{\pm}\left(x,y\right)e^{\pm i\beta_{0}z}e^{-i\omega_{0}t}\label{eq:modes1-1}\\
+\vec{E}_{pump}\left(x,y,z,t\right).\nonumber 
\end{gather}

\subsection{Material Model\label{sub:Material-Model}}

\begin{table*}
\begin{tabular}{|c|c|c|c|}
\hline 
Parameter & Description & Value $CdS$ & Value $ZnO$\tabularnewline
\hline 
\hline 
$\varepsilon_{gap,a}$ & Gap energy valence band a & $2420meV$ & $3372meV$\tabularnewline
\hline 
$\varepsilon_{gap,b}$ & Gap energy valence band b & $2435meV$ & $3382meV$\tabularnewline
\hline 
$\varepsilon_{gap,c}$ & Gap energy valence band c & - & $3416meV$\tabularnewline
\hline 
$m_{\text{eff},e}$ & Effective mass electrons & $0.1619m_{e}$ & $0.28m_{e}$\tabularnewline
\hline 
$m_{\text{eff},a}$ & Effective mass valence band a & $0.5951m_{e}$ & $0.59m_{e}$\tabularnewline
\hline 
$m_{\text{eff},b}$ & Effective mass valence band b & $0.713m_{e}$ & $0.59m_{e}$\tabularnewline
\hline 
$m_{\text{eff},c}$ & Effective mass valence band c & - & $0.45m_{e}$\tabularnewline
\hline 
$n_{bg}$ & Background refractive index & $2.81 $ & $2.0$\tabularnewline
\hline 
$d_{0}$ & Dipole matrix element & $0.279e\cdot nm$ & $0.42e\cdot nm$\tabularnewline
\hline 
$\gamma(N)$ & Polarization dephasing rate & $5ps^{-1}+4\times10^{-5}\frac{cm}{ps}N^{0.3}$ & $30ps^{-1}+2\times10^{-5}\frac{cm}{ps}N^{0.3}$\tabularnewline
\hline 
$\gamma_{rec}$ & Carrier recombination rate & $10^{9}s^{-1}$ & $10^{9}s^{-1}$\tabularnewline
\hline 
$\gamma_{f,e}$ & Intraband relaxation rate $e^{-}$ & $10^{12}s^{-1}$ & $2\times10^{12}s^{-1}$\tabularnewline
\hline 
$\gamma_{f,h}$ & Intraband relaxation rate holes & $10^{13}s^{-1}$ & $2\times10^{12}s^{-1}$\tabularnewline
\hline 
$\gamma_{ba}$ & Rel. rate valence band b to a & $6\times10^{9}s^{-1}$ & $6\times10^{9}s^{-1}$\tabularnewline
\hline 
$\gamma_{ca}$ & Rel. rate valence band c to a & - & $1\times10^{12}s^{-1}$\tabularnewline
\hline 
\end{tabular}\protect\caption{\label{tab:Material-parameters-as}Material parameters as used in
the semiconductor Bloch equations models for $ZnO$ and $CdS$.}
\end{table*}

To complete the description of the system, the polarization $\vec{P}$
has to be modelled by a separate evolution equation governed by the
material system. We use the model published in Ref. \cite{PhysRevB.91.045203},
which we shortly summarize here. Our model uses a semiconductor Bloch
equations\cite{haug2004quantum,chow1999semiconductor} approach including many-particle effects in the Screened
Hartree-Fock approximation and is adapted to 2-6 semiconductors.\cite{PhysRevB.91.045203} The
microscopic polarizations $\psi_{\lambda,q,k}$ as well as the occupation
numbers $n_{s,k}$ for conduction-band electrons ($s=e$) and for
holes in the different valence bands ($s=\lambda$) are assumed to
depend only on the absolute value $k$ of the Bloch vector $\vec{k}$. 
Then, the complex polarization in a bulk semiconductor takes the form
\begin{equation} 
\vec{P}=\sum_{\lambda,q}\int\limits_{0}^\infty\text{d}k\frac{k^{2}}{\pi^{2}}\vec{d}_{\lambda,q,k}\psi_{\lambda,q,k},
\end{equation}
where $\vec{d}_{\lambda,q,k}$ is the dipole matrix element attributed
to the transition from valence band $\lambda$ to the conduction band
and coupling to the electric field component pointing in direction
$q$. The evolution of the microscopic polarizations is described
by\cite{haug2004quantum,chow1999semiconductor} 
\begin{multline}
i\hbar\frac{\partial}{\partial t}\psi_{\lambda,q,k}=\left(1-n_{e,k}-n_{\lambda,k}\right)\Omega_{\lambda,q,k}\\
+\left(\varepsilon_{e,k}+\varepsilon_{\lambda,k}+\varepsilon_{\lambda,gap}-\triangle\varepsilon_{k}-i\gamma\left(N\right)\right)\psi_{\lambda,q,k}+\Gamma_{\psi,\lambda,q,k},\label{eq:psi}
\end{multline}
where $\Omega_{\lambda,q,k}$ are renormalized Rabi-frequencies
\begin{equation}
\Omega_{\lambda,q,k}=\vec{d}_{\lambda,q,k}\vec{E}+\int\limits_{0}^{\infty}\text{d}k'W_{k,k'}\psi_{\lambda,q,k'}.
\end{equation} The
transition energies are given by the sum of the renormalized single-particle
energies 
\begin{equation}
\varepsilon_{s,k}=\frac{\hbar^{2}k^{2}}{2m_{\text{eff},s}}-\int\limits_{0}^\infty\text{d}k'W_{k,k'}n_{s,k'},
\end{equation}
of species $s$ and the band gap of
each transition $\varepsilon_{\lambda,gap}$. In order to correctly
describe the highly excited semiconductor, the renormalized Rabi-frequencies
and transition-energies are calculated using the screened Coulomb
matrix-elements $W_{k,k'}$ instead of the unscreened matrix elements $V_{k,k'}$.\cite{haug2004quantum,PhysRevB.91.045203} Therefore the inclusion of the Coulomb-hole contribution
\begin{equation}
\triangle\varepsilon_{k}=\int\limits_{0}^\infty\text{d}k'\left(W_{k,k'}-V_{k,k'}\right)~\label{eq:chshift}
\end{equation}
 is neccessary.\cite{chow1999semiconductor} Finally, $\gamma\left(N\right)$ describes the
excitation-density dependent damping of the polarization\cite{PhysRevB.91.045203,Hugel2000,PSSB:PSSB179} and $\Gamma_{\psi,\lambda,q,k}$ represents a noise term driving spontaneous emission.\cite{Andreasen:09,PhysRevA.82.063835}

The time evolution of the occupation numbers is given by 
\begin{multline}
\frac{\partial}{\partial t}n_{e,k}=-\frac{2}{\hbar}\sum_{\lambda,q}\Im\left(\Omega_{\lambda,q,k}\psi_{\lambda,q,k}^{*}\right)-\gamma_{rec}\sum_{\lambda}n_{\lambda,k}n_{e,k}\\
+\gamma_{f,e}\left(f_{e,k}-n_{e,k}\right)+\Gamma_{n_{e},k}\label{eq:number}
\end{multline}
 for electrons and by 
\begin{multline}
\frac{\partial}{\partial t}n_{\lambda,k}=-\frac{2}{\hbar}\sum_{q}\Im\left(\Omega_{\lambda,q,k}\psi_{\lambda,q,k}^{*}\right)-\gamma_{rec}n_{\lambda,k}n_{e,k}\\
+\gamma_{f,h}\left(f_{\lambda,k}-n_{\lambda,k}\right)+\sum_{\lambda'\neq\lambda}\triangle_{\lambda\lambda'k}+\Gamma_{n_{\lambda},k}.\label{eq:number2}
\end{multline}
 for holes in the valence band $\lambda$. The first term describes
the carrier excitation by the electromagnetic field. The two terms
involving $\gamma_{\text{rec}}$ and $\gamma_{f}$ respectively describe
non-radiative recombination and intra-band relaxation of carriers
towards Fermi-Dirac distributions with a band dependent fermi level $f_{s,k}$\cite{Huang:06}. 
To allow for the relaxation between valence bands, an additional contribution $\sum_{\lambda'\neq\lambda}\triangle_{\lambda\lambda'k}$
is included in the equation of motion for the hole populations, with \begin{equation}
\triangle_{\lambda\lambda'k}=\gamma_{\lambda'\lambda}n_{\lambda',k}(1-n_{\lambda,k})-\gamma_{\lambda\lambda'}n_{\lambda,k}(1-n_{\lambda',k}).\label{eq:delta_intersubband}
\end{equation}
Similiar as in the case of the polarization, noise terms $\Gamma_{n_{e},k}$
and $\Gamma_{n_{\lambda},k}$ are added to the evolution equations
for the electron and hole occupation numbers. 

2-6 semiconductors are uniaxial crystals with the optical axis or c-axis pointing 
along the wire in $z$-direction. Optical transitions occur between a single
s-like conduction band (occupation number $n_{e,k}$) and three valence
bands (occupation numbers $\ensuremath{n_{\lambda,k}},\ensuremath{\lambda\in\left\{ a,b,c\right\} }$).\cite{PhysRev.116.573,PhysRev.128.2135} 
Assuming conservation of electron spin, allowed transitions are characterized by a conservation of angular
momenta of photons and electrons. From this, we obtain the transitions
$\psi_{a/b/c,x/y,k}$, coupling to fields polarized in the $x/y$-direction
perpendicularly to the crystals $c$-axis and the transitions $\psi_{b/c,z,k}$
coupling to the fields polarized in $z$-direction pointing along
the $c$-axis.\cite{PhysRev.116.573,PhysRev.128.2135,PhysRevB.91.045203}

The model parameters used in this publication are summarized in table
\ref{tab:Material-parameters-as}. For the verification of the model
in section \ref{sec:Verification}, we compare CMT and FDTD simulations
of an optically pumped $CdS$ nanowire laser. In accordance with the FDTD code developed in
\cite{PhysRevB.91.045203} we neglect any coupling to valence band
$c$, since the excitation is assumed to be close to the fundamental
band gap. In contrast, the third valence band is included for the simulations of $ZnO$-nanowire
lasers presented in section \ref{sec:Temporal-dynamics-of}, since the excitation frequency lies above the respective
band gap.

\subsection{Numerical Considerations\label{sub:Numerical-Considerations}}

Equation \eqref{eq:modeprop} is discretized on an equally spaced grid
in both time and propagation length
\[
\left[\hat{u}_{m}^{\pm}\right]_{j}^{i}=\hat{u}_{m}^{\pm}\left(i\triangle z,j\triangle t\right)
\]
using central finite differences with temporal and spatial discretization steps correlated as
\begin{equation}
\triangle t=\frac{\triangle z}{|v|}.\label{eq:timestep}
\end{equation}
Only for this choice numerical instabilities can be avoided. We obtain the discretized propagation equations
\begin{alignat}{1}
\left[\hat{u}_{m}^{+}\right]_{j+1}^{i+1} & =\left[\hat{u}_{m}^{+}\right]_{j}^{i}+\frac{\triangle t}{2}D_{j+\frac{1}{2}}^{i+\frac{1}{2}},\nonumber \\
\left[\hat{u}_{m}^{-}\right]_{j+1}^{i-1} & =\left[\hat{u}_{m}^{-}\right]_{j}^{i}+\frac{\triangle t}{2}D_{j+\frac{1}{2}}^{i-\frac{1}{2}}.\label{eq:timesteps}
\end{alignat}
If a material polarization as described in section \ref{sub:Material-Model}
is discretized using central finite differences, the driving terms
$D$ have to be evaluated on temporally staggered timesteps $j+\frac{1}{2}$.
As an approximation to the values required for an exact evaluation
of equation \eqref{eq:timesteps}, which are also staggered in space,
we use $D_{j+\frac{1}{2}}^{i\pm\frac{1}{2}}\approx D_{j+\frac{1}{2}}^{i}$. 

The transverse integral over the waveguide cross-section in equation
\eqref{eq:driver} can be evaluated using a spatial discretization scheme
appropriate for the waveguide geometry. Usually the transverse resolution
can be kept low, using only few data points. For instance we use a resolution
of 4 radial steps and 16 azimuthal steps for our simulations of a waveguide with a radius of $r=80nm$ presented below. 
In the case of single-mode waveguides there is no interaction between modes with different spatial profiles. Therefore the transverse resolution can in principle be restricted to a single point and the effects of the spatial mode shape can be approximated using effective values for $\gamma_m$ and $E_m$. In multimode-waveguides a higher transverse resolution has to be used, since both the coupling strength and the phase of the induced polarization have a different spatial dependence for the individual modes. For good quantitative agreement for example with the FDTD method, a transverse resolution has to be used even in the single-mode case. Therefore our model is designed to allow both for effective 1D simulations and for arbitrary transverse resolutions.

Especially for the simulation of lasing nanowires, where 
the interference between forward and backward propagating waves is strong and intensities vary on the
scale of $\frac{\lambda}{2n}$, we have to chose a fine longitudinal
discretization for the determination of the material response given by $D$. Due to the stability criterion
\eqref{eq:timestep}, this resolution requirement also determines the temporal
resolution. The material equations however are not subject to this
stabilty criterion, since they merely have to resolve the highest
frequencies present in the slowly varying envelopes. Since in our
case the computational demands are mainly determined by the solution
of the material equations, we can further reduce the time needed for
computation by restricting the number of timesteps at which the SBEs
are evaluated. This can be accomplished by skipping every $N_{skip}$
timesteps in the evaluation of the SBEs and using linear interpolation
whenever values from intermediate steps are required. In our lasing
example, we can restrict the evaluation of the material equations
to every $N_{skip}=6$th step. However this number is determined by
the detuning between material resonance and envelope center frequency
and by the shape of the envelope signals themselves and could be higher
or lower depending on the simulated scenario.

The field evolution inside the nanowire laser is heavily influenced by boundary conditions, which we will discuss next. 
Reflecting boundaries like the endfacets of a nanowire laser cavity can be implemented
by inserting the appropriate amplitudes propagating away from the
interface. Assuming that equation \eqref{eq:modes1-1} is evaluated
on points with spatial indices $i\in\left[0,N_{z}-1\right]$, general
expressions for the two missing amplitudes at the boundaries are 
\begin{alignat}{1}
\left[\hat{u}_{m}^{-}\right]_{j}^{N_{z}} & =\sum_{n}R_{mn}\left[\hat{u}_{n}^{+}\right]_{j}^{N_{z}-1}\label{eq:refl}\\
\left[\hat{u}_{m}^{+}\right]_{j}^{-1} & =\sum_{n}R_{mn}\left[\hat{u}_{n}^{-}\right]_{j}^{0}\label{eq:ref2}
\end{alignat}

The reflectivity matrix elements\cite{:/content/aip/journal/apl/83/6/10.1063/1.1599037}
$R_{mn}$ as well as the mode profiles $\vec{E}_{m}^{\pm}\left(x,y\right)$,
$\vec{H}_{m}^{\pm}\left(x,y\right)$ and the pump field $\vec{E}_{pump}\left(\vec{r},t\right)$
have to be determined beforehand by analytical estimates or with numerical
tools like FDTD using the nondispersive and linear background refractive
index of the nanowire material. 
In the unperturbed propagation equations ($D=0$), the prescribed reflectivities are the only source of reflection from the endfacets. If the mode amplitudes are coupled to the material equations, there exists an additional contribution to the reflectivity which is caused by the missing material polarization outside the simulation volume. While the prescribed reflectivities describe the reflection at the endfacet of a nanowire with the materials background index $n_{bg}$, this contribution accounts for the reflectity caused by the electronic part of the refractive index.

\section{Verification\label{sec:Verification}}

To verify our method, we compare it to direct simulations of light
propagation in semiconductor nanowires using the coupled FDTD and
semiconductor Bloch equations approach.\cite{PhysRevB.91.045203} 

\begin{figure}
\includegraphics[scale=0.55]{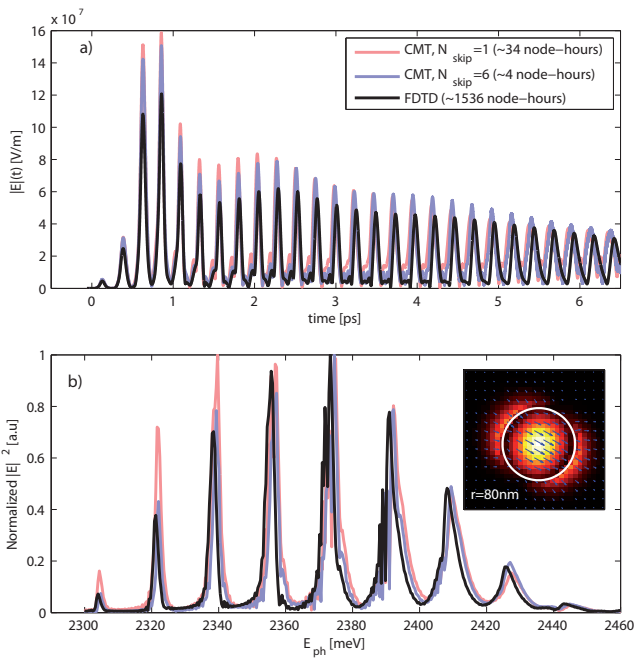}\protect\caption{\label{fig:Temporal-evolution}Temporal evolution (a) and lasing spectrum
(b) of a $CdS$ nanowire laser with radius $r=80nm$ and length $l=8\text{\textmu m}$
surrounded by air at initial quasiparticle densities of $N_{el}=3\times10^{19}cm^{-3}$,
$N_{h,a}=N_{h,b}=1.5\times10^{19}cm^{-3}$ calculated using the FDTD
method (black curve) and the CMT method using both evaluation of the
material equations at each timestep (red curves) and at each 6th timestep
(blue curves). The wire supports only the doubly degenerate fundamental
mode. The intensity profile (colour) and field direction (arrows) as
used in the FDTD simulation is shown in the insert of panel (b). The
mode is not completely rotationally symmetric, but shows lobes outside
the nanowire in polarisation direction. The mode profile used by the
CMT code is generated by interpolation of this profile onto a coarse
cylindrical coordinate system with 4 radial and 16 angular points.}
\end{figure}

Due to the initial exponential growth in intensity and the strength
of nonlinear effects, nanowire lasers constitute an extremely demanding
test case for our model. We now consider a highly excited
$CdS$ ($n_{bg}=2.81$) nanowire laser with length $l=8\mu m$ and
radius $r=80nm$. At a central wavelength of $\lambda=512nm$, the
wire only supports the two degenerate fundamental modes as shown in
the inset of Fig. \ref{fig:Temporal-evolution}(b)), propagating
with group velocity $v_{g}=7.72\times10^{7}\frac{m}{s}$ and propagation
constant $\beta_{0}=22.28\frac{1}{\mu m}$. To create a laser cavity,
the wire is terminated by realistic endfacets with an amplitude reflectivity
matrix as used in equations \eqref{eq:refl} and \eqref{eq:ref2} 
\begin{equation}
R=\left(\begin{array}{cc}
0.51 & 0.02\\
0.02 & 0.51
\end{array}\right)
\end{equation}
 extracted from a simple linear FDTD simulation of a wire endfacet\cite{:/content/aip/journal/apl/83/6/10.1063/1.1599037}.
The occupation probabilities for electrons and holes are initialized
with Fermi-distributions at $T=300K$ for spatial densities of $N_{el}=3\times10^{19}cm^{-3}$,
$N_{h,a}=N_{h,b}=1.5\times10^{19}cm^{-3}$. In order to avoid a randomization
of the lasing output which would render direct comparisons between
the two codes difficult, we switch off spontaneous emission and instead
start the laser with a sech-shaped seed pulse with a pulse duration of $w_t=20fs$. The spatial resolution in the FDTD
simulation is $\triangle x=\triangle y=\triangle z=10nm$. In the
CMT simulation we use a longitudinal resolution of $\triangle z=10nm$,
while the transverse fields are sampled using a cylindrical coordinate
system with $N_{R}=4$ radial steps.
Due to the high mode confinement, the modal fields have a significant longitudinal contribution,
which is not radially symmetric. Therefore, also an azimuthal resolution with $N_{\theta}=16$ steps is neccessary
in order to achieve good agreement with FDTD simulations.
We use both simulations where the material equations are solved on
the same temporal grid as the propagation equations ($N_{skip}=1$)
and on a grid with $N_{skip}=6$. In this special case the simulation
becomes unstable, if we try to further reduce the temporal resolution
of the material equations. However the CMT variety with $N_{skip}=6$
already is approximately 400 times as fast as the FDTD method. 

Fig. \ref{fig:Temporal-evolution}(a) shows the time-resolved electrical field strength recorded at one endfacet of
the nanowire. Since the seed pulse is amplified and reflected inside the cavity, a train of pulses is emitted at the endfacet. We initially observe a fast rise in lasing intensity.
After the inversion in the spectral region of the main longitudinal
mode has been depleted, the emission switches to other longitudinal
modes with lower gain, leading to a slow decay in the lasing intensity
similiar to the results in \cite{PhysRevB.91.045203}. Despite the
dramatically reduced computation time, the CMT results show excellent
agreement with the FDTD results concerning the spectral position of
the lasing modes(Fig. \ref{fig:Temporal-evolution}(b)). The temporal
dynamics (Fig. \ref{fig:Temporal-evolution}(a)) also show good qualitative
agreement concerning the shape of the emitted pulse train as well
as the position of the individual pulses, which is directly linked
to the modal dispersion under the influence of the material system.
However, the overall intensity of the emitted pulse train is higher
in the CMT simulation. Near-perfect agreement can be achieved by tuning
of the endfacet reflectivity used by the CMT code. From this we conclude,
that the difference in emission intensity is caused by the imperfect
modelling of the endfacet reflectivity under the influence of the
material system, which is an inherent limitation of Coupled Mode Theory.
However, the endfacet reflectivities of real nanowires will always vary due to imperfections 
in growth and preparation, giving rise to fluctuations of laser performance in experiments.
Most importantly, the essential laser dynamics are already captured very well
by our model.

\section{Temporal dynamics of Nanowire Lasers\label{sec:Temporal-dynamics-of}}

\begin{figure}
\includegraphics[scale=0.65]{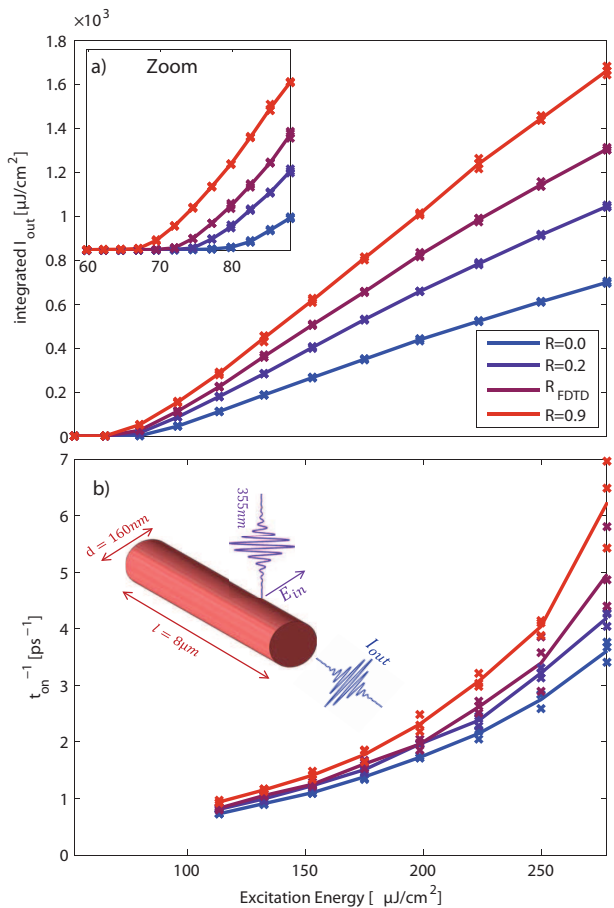}

\protect\caption{\label{fig:(a):-Lasing-curve} Emission properties of exemplary $ZnO$-nanowire
lasers with different enfacet reflectivities (a): Output energy for
different pump energies, the inset shows a zoom into the region around
the laser-threshold. (b): Inverse emission onset time for different
pump energies in the lasing regime. The results from individual runs
with different random number seeds for the spontaneous emission noise
are shown by markers. The lineplots show the averaged data for each
nanowire configuration.}
\end{figure}
\begin{figure}
\includegraphics[scale=0.65]{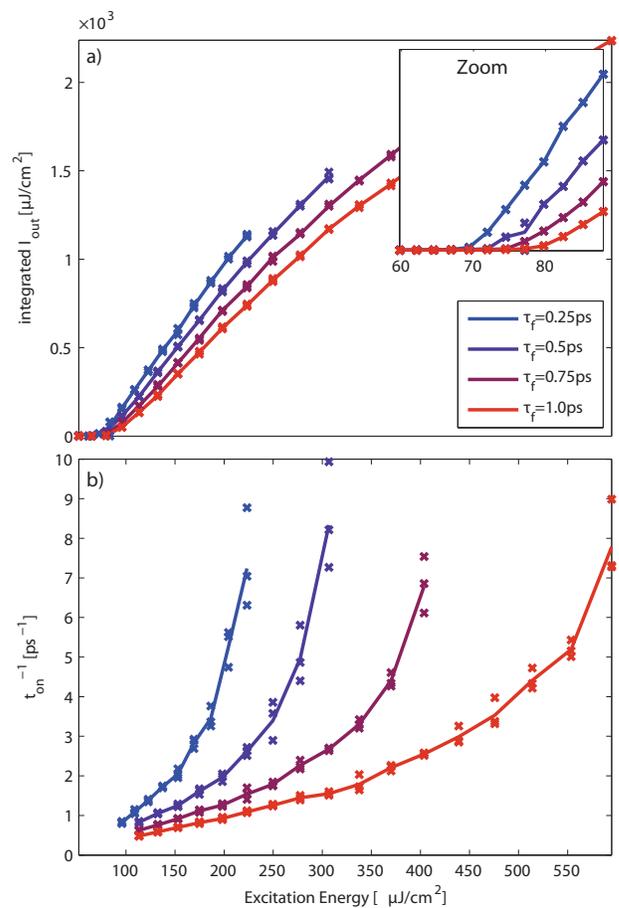}

\protect\caption{\label{fig:(a):-Lasing-curve-1}Emission properties of exemplary $ZnO$-nanowire
lasers with different intraband relaxation times $\tau_{f}=\frac{1}{\gamma_{f}}$
(a): Output energy for different pump energies. (b): Inverse emission
onset time for different pump energies in the lasing regime. The solid
lines show data averaged over three runs with different random seeds
(crosses).}
\end{figure}
Since the increased efficiency of the presented model allows us to
simulate the entire excitation and lasing process in a comparatively
short time, it is now possible to perform a more comprehensive analysis
of the properties of semiconductor nanowire lasers. In our further analysis, we choose $ZnO$
wires ($n_{bg}=2.0$) due to the wealth of available experimental
data\cite{0957-4484-27-22-225702,doi:10.1021/acs.nanolett.5b01271,doi:10.1021/acs.nanolett.6b00811}
for this material system. We specifically model our nanowires after
experiments performed by Wille et al. \cite{0957-4484-27-22-225702},
where time-resolved $\mu$-PL measurements have been used in order
to investigate the temporal dynamics of the lasing emission. An important
parameter extracted from the experiment is the emission onset time
$t_{on}$, which is defined as the time between the maximum of the
pump pulse and the buildup of the laser emission to $\frac{1}{e}$
of its maximum. In the lasing regime, the inverse emission onset time
$t_{on}^{-1}$ is reported to increase nonlinearly with increasing
pump power. Additionally, time-resolved spectra have been obtained
from the experiment and a spectral red-shift of the lasing modes occuring
during emission has been observed. This has been attributed to the
depletion of excited carriers due to stimulated emission, which leads
to an increase of the refractive index and therefore to a shift in
the energies of the wires longitudinal Fabry-Perot modes.

While the main goal of our simulations is the reproduction of the experimental
features outlined above, we are additionally going to investigate
the dependence of the laser emission on parameters which can vary
from sample to sample and are not easily accessible in experiments. We study
exemplary nanowires with a length $l=8\mu m$ and a diameter of $d=160nm$
as sketched in the inset of Fig. \ref{fig:(a):-Lasing-curve}(b).
Since the wire diameter is definitely in the single-mode
regime, we restrict our simulations to the fundamental $HE01$ mode with a propagation
constant $\beta_{0}=24.32\frac{1}{\mu m}$ and group velocity $v_{g}=1.26\times10^{8}\frac{m}{s}$.
The diagonal endfacet reflectivity $R_{m,m}=0.29$ extracted from
FDTD simulations is slightly lower than for the $CdS$-wire. To decrease
simulation time, we restrict the transverse resolution to a single
point. This will not exactly reproduce results as they would be obtained
by an FDTD-simulation, but allows us to capture the essential
physics in the present case of a single-mode wire.
The wires are optically pumped from above with sech-shaped pulses with a temporal width of $w_{t}=2ps$
and a central wavelength of $\lambda=355nm$, polarized perpendicularly
to the wire axis. The exciting field is assumed to be homogeneous along the wire.
Since the output can vary from realization to realization due to different random seed
values for the spontaneous emission noise in equations \eqref{eq:psi},
\eqref{eq:number} and \eqref{eq:number2}, we average over several simulation
runs. As the fluctuations are not excessively strong, the averaging
procedure is restricted to three runs. 

First, we investigate the emission properties of several wires with
varying endfacet reflectivity in the range between the extreme values
$R=0.0$ and $R=0.9$ and including the reflectivity $R_{FDTD}$ as
extracted from FDTD simulations. The endfacet reflectivity of real
nanowires can vary strongly between different samples and strongly
affects the quality of the optical cavity. Note, that simulations
with a nominal endfacet reflectivity of $R=0.0$ still have a finite
reflectivity due to the material boundary effect described in section
\ref{sub:Numerical-Considerations}. The input/output curves resulting
from our simulations are given in Fig.\ref{fig:(a):-Lasing-curve}(a)
and show the typical lasing behaviour which exhibits a linear increase
above a certain threshold intensity of approximately $W_{th}=70\mu J/cm^2$. The laser threshold power lies below the experimental value of $W_{th}=200\mu J/cm^2$, but is of the same order of magnitude. This is to be expected, since we prescribe a homogeneous electric field inside the wire and do not take into account effects like scattering of the exciting waves from the wire or a spatially inhomogeneous excitation.
As expected, wires with lower endfacet reflectivities achieve a lower efficiency. Fig.\ref{fig:(a):-Lasing-curve}(b)
shows the inverse emission onset time $t_{on}^{-1}$. As reported
in the experiment, lasing generally sets in faster with increasing
excitation power. We also observe an increase of $t_{on}^{-1}$ for
higher endfacet reflectivities. Since our model does not include an
excitation-dependent carrier relaxation time, this increase can be
explained solely by the fact that the laser threshold is reached at
an earlier time, if a stronger pump pulse is used. The same effect
occurs, if we lower the lasing threshold by increasing the endfacet
reflectivity (see inset of Fig. \ref{fig:(a):-Lasing-curve}(a)).
Additionally, roundtrip losses are lowered in this case, allowing
for a faster buildup of lasing oscillations.

After having investigated the influence of the optical cavity by simulating
wires with different endfacet reflectivities, we now consider the influence
of the semiconductor relaxation dynamics, which are governed by the
intraband relaxation time $\tau_{f}=\frac{1}{\gamma}_{f}$. 

Fig. \ref{fig:(a):-Lasing-curve-1}(a) shows simulated input/output
curves for different values of $\tau_{f}$. We observe a decreasing
lasing efficiency for increasing intraband relaxation times. This is due to
the fact, that the lasing process is slowed down for high relaxation
times. Thus, a higher number of carriers can recombine by the way
of other relaxation processes before being used for spontaneous emission.
As expected, the slowed down dynamics also leads to a strong decrease
in the inverse emission onset time $t_{on}^{-1}$ (\ref{fig:(a):-Lasing-curve-1}(b)).
However, the nonlinear increase of $t_{on}^{-1}$ with increasing
pump power is retained for all simulated values of $\tau_{f}$. 

\begin{figure}
\includegraphics[scale=0.55]{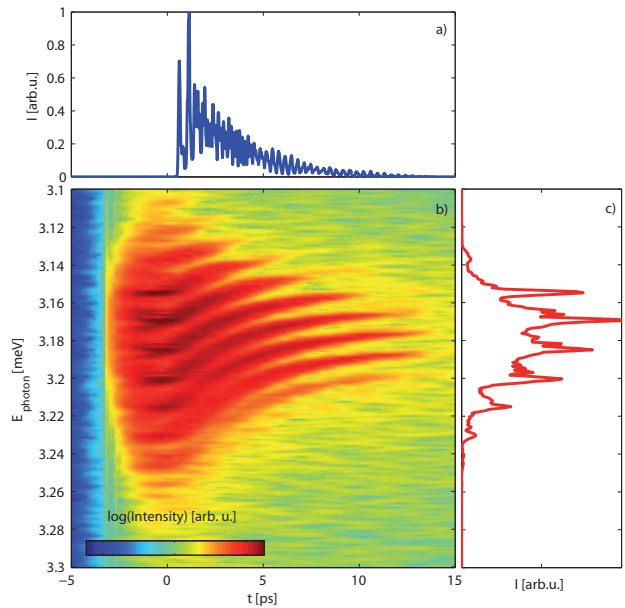}
\protect\caption{\label{fig:(a):-Temporal-intensity}(a): Temporal intensity profile
of lasing output from an exemplary $ZnO$ nanowire. (b): Dynamics
of the individual lasing modes obtained from a windowed Fourier-transform
of the temporal data. (c): Spectral intensity of the laser emission
obtained from a Fourier-transform over the whole output pulse duration.
All data has been averaged over three simulation runs with different
random number input for the spontaneous emission. }
\end{figure}

Finally, we use a windowed Fourier-transform in order to obtain the
time-resolved lasing spectrum (Fig. \ref{fig:(a):-Temporal-intensity}(b))
of one of our wire configurations with $R=R_{FDTD}$ and $\tau_{f}=0.5ps$, 
which produces similiar temporal dynamics as measured in the experiment.\cite{0957-4484-27-22-225702}

The temporal lasing profile and the time-averaged spectrum are given
in (Fig. \ref{fig:(a):-Temporal-intensity}) (a) and (c), respectively.
Similiar to the experiment, we observe a pronounced red-shift of the
lasing modes during the emission process due to the change in the
materials refractive index caused by the depletion of quasi-particles.

\begin{figure}
\includegraphics[scale=0.65]{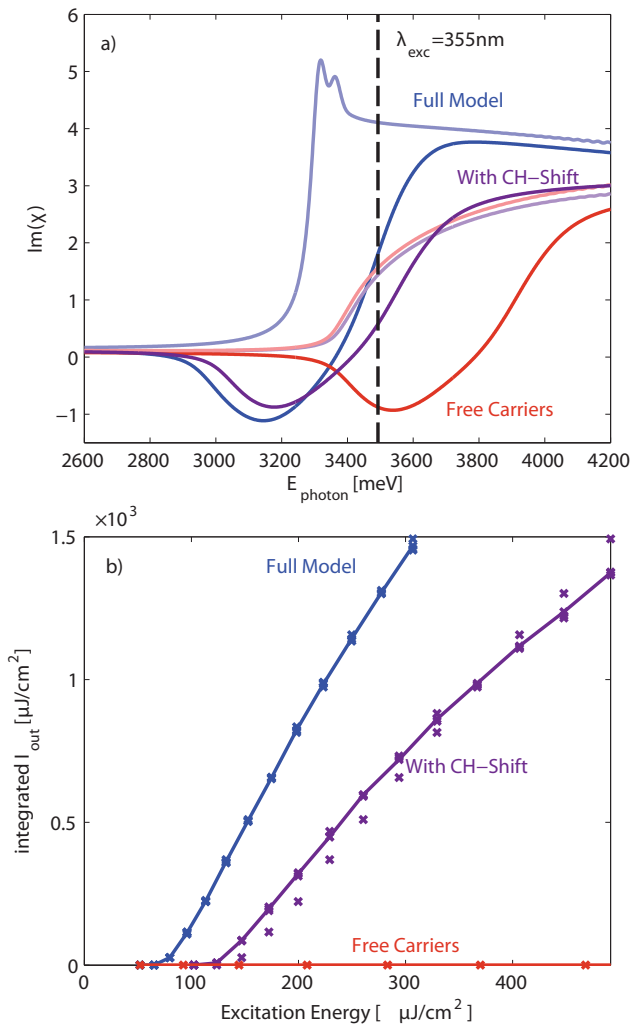}   
\protect\caption{\label{fig:LasingCouNoCou} Influence of the Coulomb interaction on the laser properties of a $ZnO$-nanowire pumped at $\lambda_{exc}=355nm$. (a): Absorption spectra for the full model (blue), the free carrier model without the Coulomb interaction (red) and a free carrier model, where the Coulomb-hole shift is artificially included in order to correct the spectral position of the gain region (violet). Spectra are plotted both at low excitation ($N=3\times10^{10}cm^{-3}$ , faded lines) and in the gain regime ($N=3\times10^{19}cm^{-3}$, strong lines). The dashed line shows the excitation wavelength. (b): Corresponding lasing curves. The line coulours correspond to the three models from panel (a).}
\end{figure}

Since the most computationally demanding part of our model is the evaluation of the Coulomb terms, it would be desirable to simplify the model in this regard. Therefore we also examine the influence of the Coulomb interaction on the laser performance of our exemplary $ZnO$-nanowire (Fig. \ref{fig:LasingCouNoCou}).
Several changes can already be predicted by considering the differences in the linear absorption and gain spectra for the full model (blue lines in Fig. \ref{fig:LasingCouNoCou}(a)) and the free-carrier model (red lines) at low excitation ($N=3\times10^{10}cm^{-3}$ , faded lines) and at high excitation ($N=3\times10^{19}cm^{-3}$, strong lines). First, the Coulomb interaction leads to a well-known enhancement of the overall optical response\cite{haug2004quantum,chow1999semiconductor}. If the Coulomb interaction is omitted, we expect a decrease in absorption at the pump wavelength, leading to a higher laser threshold and a decreased laser efficiency. Second, the omission of the various Coulomb-related shifts in the quasi-particle energies will lead to a significant blue-shift of the gain region, which now will be positioned almost exclusively at energies above the bandgap. It is possible to artificially correct this by including the Coulomb-hole shift $\triangle\varepsilon_{k}$ (Eq. \eqref{eq:chshift}) into the free carrier model. Even though this is physically not meaningful in the context of an interaction-free model, the position of the gain region can be partly corrected in this way (violet curves in Fig. \ref{fig:LasingCouNoCou}). 
The corresponding lasing curves are shown in Fig. \ref{fig:LasingCouNoCou}(b). As it turns out, the decreased density of states and the lack of Coulomb-related energy shifts in the free-carrier model(red curve) completely obstruct optical pumping up to the laser threshold for a pump-wavelength of $\lambda_{exc}=355nm$, since the relevant transitions saturate before the neccessary excitation density is reached. For the artificially corrected model with the Coulomb-hole shift included (violet curve), we observe lasing. However as we would expect, we observe an increased laser threshold and a decreased efficiency as compared to the full model(blue curve). We conclude, that an omission of the Coulomb interaction terms leads to significant differences in the predicted performance and spectral properties of semiconductor lasers. A thus reduced model requires significant tuning of parameters in order to yield meaningful results.

\section{Conclusion}

In conclusion, we have developed a new theoretical model for the simulation
of light-matter interaction and lasing in semiconductor nanowire structures
based on the framework of coupled mode theory coupled to semiconductor Bloch equations. 
We have shown that our model can qualitatively and quantitatively reproduce the results
achieved by a more general FDTD model, but with a speedup of up to
three orders of magnitude depending on the simulated setup and the required accuracy. 
We have further applied the model to the simulation of the properties
and temporal dynamics of $ZnO$-nanowire lasers. We reproduce
the red-shift of the lasing modes occuring during the emission process
as well as the nonlinear increase of the inverse emission onset time
with increasing excitation power, which have been observed in experiments\cite{0957-4484-27-22-225702}.
Further, we have studied the influence of the optical cavity as well
as the carrier relaxation time on the laser dynamics. We observe a reduction of the 
emission onset time both with increasing endfacet
reflectivity and with decreasing material relaxation time. In the
future, the computational efficiency of our model is going to pave
the way for the inclusion of more advanced material models as well
as more complicated geometries in full time-domain simulations of
semiconductor nanowires.
\begin{acknowledgments}
The authors gratefully acknowledge financial support by Deutsche Forschungsgemeinschaft
(Forschergruppe FOR1616, projects P5 and E4). The authors thank Robert
Röder and Marcel Wille for helpful discussions.
\end{acknowledgments}


\begin{thebibliography}{28}%
\makeatletter
\providecommand \@ifxundefined [1]{%
 \@ifx{#1\undefined}
}%
\providecommand \@ifnum [1]{%
 \ifnum #1\expandafter \@firstoftwo
 \else \expandafter \@secondoftwo
 \fi
}%
\providecommand \@ifx [1]{%
 \ifx #1\expandafter \@firstoftwo
 \else \expandafter \@secondoftwo
 \fi
}%
\providecommand \natexlab [1]{#1}%
\providecommand \enquote  [1]{``#1''}%
\providecommand \bibnamefont  [1]{#1}%
\providecommand \bibfnamefont [1]{#1}%
\providecommand \citenamefont [1]{#1}%
\providecommand \href@noop [0]{\@secondoftwo}%
\providecommand \href [0]{\begingroup \@sanitize@url \@href}%
\providecommand \@href[1]{\@@startlink{#1}\@@href}%
\providecommand \@@href[1]{\endgroup#1\@@endlink}%
\providecommand \@sanitize@url [0]{\catcode `\\12\catcode `\$12\catcode
  `\&12\catcode `\#12\catcode `\^12\catcode `\_12\catcode `\%12\relax}%
\providecommand \@@startlink[1]{}%
\providecommand \@@endlink[0]{}%
\providecommand \url  [0]{\begingroup\@sanitize@url \@url }%
\providecommand \@url [1]{\endgroup\@href {#1}{\urlprefix }}%
\providecommand \urlprefix  [0]{URL }%
\providecommand \Eprint [0]{\href }%
\providecommand \doibase [0]{http://dx.doi.org/}%
\providecommand \selectlanguage [0]{\@gobble}%
\providecommand \bibinfo  [0]{\@secondoftwo}%
\providecommand \bibfield  [0]{\@secondoftwo}%
\providecommand \translation [1]{[#1]}%
\providecommand \BibitemOpen [0]{}%
\providecommand \bibitemStop [0]{}%
\providecommand \bibitemNoStop [0]{.\EOS\space}%
\providecommand \EOS [0]{\spacefactor3000\relax}%
\providecommand \BibitemShut  [1]{\csname bibitem#1\endcsname}%
\let\auto@bib@innerbib\@empty
\bibitem [{\citenamefont {Duan}\ \emph {et~al.}(2003)\citenamefont {Duan},
  \citenamefont {Huang}, \citenamefont {Agarwal},\ and\ \citenamefont
  {Lieber}}]{duansingle2003}%
  \BibitemOpen
  \bibfield  {author} {\bibinfo {author} {\bibfnamefont {Xiangfeng}\
  \bibnamefont {Duan}}, \bibinfo {author} {\bibfnamefont {Yu}~\bibnamefont
  {Huang}}, \bibinfo {author} {\bibfnamefont {Ritesh}\ \bibnamefont {Agarwal}},
  \ and\ \bibinfo {author} {\bibfnamefont {Charles~M.}\ \bibnamefont
  {Lieber}},\ }\bibfield  {title} {\enquote {\bibinfo {title} {Single-nanowire
  electrically driven lasers},}\ }\href {\doibase 10.1038/nature01353}
  {\bibfield  {journal} {\bibinfo  {journal} {Nature}\ }\textbf {\bibinfo
  {volume} {421}},\ \bibinfo {pages} {241--245} (\bibinfo {year}
  {2003})}\BibitemShut {NoStop}%
\bibitem [{\citenamefont {Oulton}\ \emph {et~al.}(2009)\citenamefont {Oulton},
  \citenamefont {Sorger}, \citenamefont {Zentgraf}, \citenamefont {Ma},
  \citenamefont {Gladden}, \citenamefont {Dai}, \citenamefont {Bartal},\ and\
  \citenamefont {Zhang}}]{oultonplasmon2009}%
  \BibitemOpen
  \bibfield  {author} {\bibinfo {author} {\bibfnamefont {Rupert~F.}\
  \bibnamefont {Oulton}}, \bibinfo {author} {\bibfnamefont {Volker~J.}\
  \bibnamefont {Sorger}}, \bibinfo {author} {\bibfnamefont {Thomas}\
  \bibnamefont {Zentgraf}}, \bibinfo {author} {\bibfnamefont {Ren-Min}\
  \bibnamefont {Ma}}, \bibinfo {author} {\bibfnamefont {Christopher}\
  \bibnamefont {Gladden}}, \bibinfo {author} {\bibfnamefont {Lun}\ \bibnamefont
  {Dai}}, \bibinfo {author} {\bibfnamefont {Guy}\ \bibnamefont {Bartal}}, \
  and\ \bibinfo {author} {\bibfnamefont {Xiang}\ \bibnamefont {Zhang}},\
  }\bibfield  {title} {\enquote {\bibinfo {title} {Plasmon lasers at deep
  subwavelength scale},}\ }\href {\doibase 10.1038/nature08364} {\bibfield
  {journal} {\bibinfo  {journal} {Nature}\ }\textbf {\bibinfo {volume} {461}},\
  \bibinfo {pages} {629--632} (\bibinfo {year} {2009})}\BibitemShut {NoStop}%
\bibitem [{\citenamefont {Sidiropoulos}\ \emph {et~al.}(2014)\citenamefont
  {Sidiropoulos}, \citenamefont {R\"{o}der}, \citenamefont {Geburt},
  \citenamefont {Hess}, \citenamefont {Maier}, \citenamefont {Ronning},\ and\
  \citenamefont {Oulton}}]{Sidiropoulos2014}%
  \BibitemOpen
  \bibfield  {author} {\bibinfo {author} {\bibfnamefont {Themistoklis P.~H.}\
  \bibnamefont {Sidiropoulos}}, \bibinfo {author} {\bibfnamefont {Robert}\
  \bibnamefont {R\"{o}der}}, \bibinfo {author} {\bibfnamefont {Sebastian}\
  \bibnamefont {Geburt}}, \bibinfo {author} {\bibfnamefont {Ortwin}\
  \bibnamefont {Hess}}, \bibinfo {author} {\bibfnamefont {Stefan~A.}\
  \bibnamefont {Maier}}, \bibinfo {author} {\bibfnamefont {Carsten}\
  \bibnamefont {Ronning}}, \ and\ \bibinfo {author} {\bibfnamefont {Rupert~F.}\
  \bibnamefont {Oulton}},\ }\bibfield  {title} {\enquote {\bibinfo {title}
  {Ultrafast plasmonic nanowire lasers near the surface plasmon frequency},}\
  }\href {\doibase 10.1038/nphys3103} {\bibfield  {journal} {\bibinfo
  {journal} {Nature Physics}\ }\textbf {\bibinfo {volume} {10}},\ \bibinfo
  {pages} {870--876} (\bibinfo {year} {2014})}\BibitemShut {NoStop}%
\bibitem [{\citenamefont {Saxena}\ \emph {et~al.}(2013)\citenamefont {Saxena},
  \citenamefont {Mokkapati}, \citenamefont {Parkinson}, \citenamefont {Jiang},
  \citenamefont {Gao}, \citenamefont {Tan},\ and\ \citenamefont
  {Jagadish}}]{saxenaoptically2013}%
  \BibitemOpen
  \bibfield  {author} {\bibinfo {author} {\bibfnamefont {Dhruv}\ \bibnamefont
  {Saxena}}, \bibinfo {author} {\bibfnamefont {Sudha}\ \bibnamefont
  {Mokkapati}}, \bibinfo {author} {\bibfnamefont {Patrick}\ \bibnamefont
  {Parkinson}}, \bibinfo {author} {\bibfnamefont {Nian}\ \bibnamefont {Jiang}},
  \bibinfo {author} {\bibfnamefont {Qiang}\ \bibnamefont {Gao}}, \bibinfo
  {author} {\bibfnamefont {Hark~Hoe}\ \bibnamefont {Tan}}, \ and\ \bibinfo
  {author} {\bibfnamefont {Chennupati}\ \bibnamefont {Jagadish}},\ }\bibfield
  {title} {\enquote {\bibinfo {title} {Optically pumped room-temperature gaas
  nanowire lasers},}\ }\href {\doibase 10.1038/nphoton.2013.303} {\bibfield
  {journal} {\bibinfo  {journal} {Nature Photonics}\ }\textbf {\bibinfo
  {volume} {7}},\ \bibinfo {pages} {963--968} (\bibinfo {year}
  {2013})}\BibitemShut {NoStop}%
\bibitem [{\citenamefont {Röder}\ \emph {et~al.}(2013)\citenamefont {Röder},
  \citenamefont {Wille}, \citenamefont {Geburt}, \citenamefont {Rensberg},
  \citenamefont {Zhang}, \citenamefont {Lu}, \citenamefont {Capasso},
  \citenamefont {Buschlinger}, \citenamefont {Peschel},\ and\ \citenamefont
  {Ronning}}]{doi:10.1021/nl401355b}%
  \BibitemOpen
  \bibfield  {author} {\bibinfo {author} {\bibfnamefont {Robert}\ \bibnamefont
  {Röder}}, \bibinfo {author} {\bibfnamefont {Marcel}\ \bibnamefont {Wille}},
  \bibinfo {author} {\bibfnamefont {Sebastian}\ \bibnamefont {Geburt}},
  \bibinfo {author} {\bibfnamefont {Jura}\ \bibnamefont {Rensberg}}, \bibinfo
  {author} {\bibfnamefont {Mengyao}\ \bibnamefont {Zhang}}, \bibinfo {author}
  {\bibfnamefont {Jia~Grace}\ \bibnamefont {Lu}}, \bibinfo {author}
  {\bibfnamefont {Federico}\ \bibnamefont {Capasso}}, \bibinfo {author}
  {\bibfnamefont {Robert}\ \bibnamefont {Buschlinger}}, \bibinfo {author}
  {\bibfnamefont {Ulf}\ \bibnamefont {Peschel}}, \ and\ \bibinfo {author}
  {\bibfnamefont {Carsten}\ \bibnamefont {Ronning}},\ }\bibfield  {title}
  {\enquote {\bibinfo {title} {Continuous wave nanowire lasing},}\ }\href
  {\doibase 10.1021/nl401355b} {\bibfield  {journal} {\bibinfo  {journal} {Nano
  Letters}\ }\textbf {\bibinfo {volume} {13}},\ \bibinfo {pages} {3602--3606}
  (\bibinfo {year} {2013})},\ \Eprint
  {http://arxiv.org/abs/http://pubs.acs.org/doi/pdf/10.1021/nl401355b}
  {http://pubs.acs.org/doi/pdf/10.1021/nl401355b} \BibitemShut {NoStop}%
\bibitem [{\citenamefont {ElSayed}\ \emph {et~al.}(1994)\citenamefont
  {ElSayed}, \citenamefont {B{\'a}nyai},\ and\ \citenamefont
  {Haug}}]{ElSayed:94b}%
  \BibitemOpen
  \bibfield  {author} {\bibinfo {author} {\bibfnamefont {K.}~\bibnamefont
  {ElSayed}}, \bibinfo {author} {\bibfnamefont {L.}~\bibnamefont {B{\'a}nyai}},
  \ and\ \bibinfo {author} {\bibfnamefont {H.}~\bibnamefont {Haug}},\
  }\bibfield  {title} {\enquote {\bibinfo {title} {Coulomb quantum kinetics and
  optical dephasing on the femtosecond time scale},}\ }\href@noop {} {\bibfield
   {journal} {\bibinfo  {journal} {Phys. Rev. B}\ }\textbf {\bibinfo {volume}
  {\textbf{50}}},\ \bibinfo {pages} {1541} (\bibinfo {year}
  {1994})}\BibitemShut {NoStop}%
\bibitem [{\citenamefont {Manzke}\ and\ \citenamefont
  {Henneberger}(2002)}]{Manzke:02}%
  \BibitemOpen
  \bibfield  {author} {\bibinfo {author} {\bibfnamefont {G.}~\bibnamefont
  {Manzke}}\ and\ \bibinfo {author} {\bibfnamefont {K.}~\bibnamefont
  {Henneberger}},\ }\bibfield  {title} {\enquote {\bibinfo {title}
  {Quantum-kinetic effects in the linear optical response of {GaAs} quantum
  wells},}\ }\href@noop {} {\bibfield  {journal} {\bibinfo  {journal} {phys.
  stat. sol. (b)}\ }\textbf {\bibinfo {volume} {\textbf{234}}},\ \bibinfo
  {pages} {233} (\bibinfo {year} {2002})}\BibitemShut {NoStop}%
\bibitem [{\citenamefont {Chow}\ and\ \citenamefont
  {Koch}(1999)}]{chow1999semiconductor}%
  \BibitemOpen
  \bibfield  {author} {\bibinfo {author} {\bibfnamefont {W.W.}\ \bibnamefont
  {Chow}}\ and\ \bibinfo {author} {\bibfnamefont {S.W.}\ \bibnamefont {Koch}},\
  }\href {http://books.google.de/books?id=ltEAFxGT3pgC} {\emph {\bibinfo
  {title} {Semiconductor-Laser Fundamentals: Physics of the Gain Materials}}}\
  (\bibinfo  {publisher} {Springer},\ \bibinfo {year} {1999})\BibitemShut
  {NoStop}%
\bibitem [{\citenamefont {Haug}\ and\ \citenamefont
  {Koch}(2004)}]{haug2004quantum}%
  \BibitemOpen
  \bibfield  {author} {\bibinfo {author} {\bibfnamefont {H.}~\bibnamefont
  {Haug}}\ and\ \bibinfo {author} {\bibfnamefont {S.W.}\ \bibnamefont {Koch}},\
  }\href {http://books.google.de/books?id=-UoG0Hx0w04C} {\emph {\bibinfo
  {title} {Quantum Theory of the Optical and Electronic Properties of
  Semiconductors (4th Edition)}}}\ (\bibinfo  {publisher} {World Scientific},\
  \bibinfo {year} {2004})\BibitemShut {NoStop}%
\bibitem [{\citenamefont {Yee}(1966)}]{Yee66numericalsolution}%
  \BibitemOpen
  \bibfield  {author} {\bibinfo {author} {\bibfnamefont {Kane~S.}\ \bibnamefont
  {Yee}},\ }\bibfield  {title} {\enquote {\bibinfo {title} {Numerical solution
  of initial boundary value problems involving maxwells equations in isotropic
  media},}\ }\href@noop {} {\bibfield  {journal} {\bibinfo  {journal} {IEEE
  Trans. Antennas and Propagation}\ ,\ \bibinfo {pages} {302--307}} (\bibinfo
  {year} {1966})}\BibitemShut {NoStop}%
\bibitem [{\citenamefont {Taflove}\ and\ \citenamefont
  {Hagness}(2005)}]{taflove:2005}%
  \BibitemOpen
  \bibfield  {author} {\bibinfo {author} {\bibfnamefont {Allen}\ \bibnamefont
  {Taflove}}\ and\ \bibinfo {author} {\bibfnamefont {Susan~C.}\ \bibnamefont
  {Hagness}},\ }\href {http://www.worldcat.org/isbn/1580538320} {\emph
  {\bibinfo {title} {Computational Electrodynamics: The Finite-Difference
  Time-Domain Method, Third Edition}}},\ \bibinfo {edition} {3rd}\ ed.\
  (\bibinfo  {publisher} {Artech House},\ \bibinfo {year} {2005})\BibitemShut
  {NoStop}%
\bibitem [{\citenamefont {Buschlinger}\ \emph {et~al.}(2015)\citenamefont
  {Buschlinger}, \citenamefont {Lorke},\ and\ \citenamefont
  {Peschel}}]{PhysRevB.91.045203}%
  \BibitemOpen
  \bibfield  {author} {\bibinfo {author} {\bibfnamefont {Robert}\ \bibnamefont
  {Buschlinger}}, \bibinfo {author} {\bibfnamefont {Michael}\ \bibnamefont
  {Lorke}}, \ and\ \bibinfo {author} {\bibfnamefont {Ulf}\ \bibnamefont
  {Peschel}},\ }\bibfield  {title} {\enquote {\bibinfo {title} {Light-matter
  interaction and lasing in semiconductor nanowires: A combined
  finite-difference time-domain and semiconductor bloch equation approach},}\
  }\href {\doibase 10.1103/PhysRevB.91.045203} {\bibfield  {journal} {\bibinfo
  {journal} {Phys. Rev. B}\ }\textbf {\bibinfo {volume} {91}},\ \bibinfo
  {pages} {045203} (\bibinfo {year} {2015})}\BibitemShut {NoStop}%
\bibitem [{\citenamefont {Guazzotti}\ \emph {et~al.}(2016)\citenamefont
  {Guazzotti}, \citenamefont {Pusch}, \citenamefont {Reiter},\ and\
  \citenamefont {Hess}}]{PhysRevB.94.115303}%
  \BibitemOpen
  \bibfield  {author} {\bibinfo {author} {\bibfnamefont {Stefano}\ \bibnamefont
  {Guazzotti}}, \bibinfo {author} {\bibfnamefont {Andreas}\ \bibnamefont
  {Pusch}}, \bibinfo {author} {\bibfnamefont {Doris~E.}\ \bibnamefont
  {Reiter}}, \ and\ \bibinfo {author} {\bibfnamefont {Ortwin}\ \bibnamefont
  {Hess}},\ }\bibfield  {title} {\enquote {\bibinfo {title} {Dynamical
  calculation of third-harmonic generation in a semiconductor quantum well},}\
  }\href {\doibase 10.1103/PhysRevB.94.115303} {\bibfield  {journal} {\bibinfo
  {journal} {Phys. Rev. B}\ }\textbf {\bibinfo {volume} {94}},\ \bibinfo
  {pages} {115303} (\bibinfo {year} {2016})}\BibitemShut {NoStop}%
\bibitem [{\citenamefont {B{\"o}hringer}\ and\ \citenamefont
  {Hess}(2008{\natexlab{a}})}]{Boehringer2008159}%
  \BibitemOpen
  \bibfield  {author} {\bibinfo {author} {\bibfnamefont {Klaus}\ \bibnamefont
  {B{\"o}hringer}}\ and\ \bibinfo {author} {\bibfnamefont {Ortwin}\
  \bibnamefont {Hess}},\ }\bibfield  {title} {\enquote {\bibinfo {title} {A
  full-time-domain approach to spatio-temporal dynamics of semiconductor
  lasers. i. theoretical formulation},}\ }\href {\doibase
  http://dx.doi.org/10.1016/j.pquantelec.2008.10.002} {\bibfield  {journal}
  {\bibinfo  {journal} {Progress in Quantum Electronics}\ }\textbf {\bibinfo
  {volume} {32}},\ \bibinfo {pages} {159--246} (\bibinfo {year}
  {2008}{\natexlab{a}})}\BibitemShut {NoStop}%
\bibitem [{\citenamefont {B{\"o}hringer}\ and\ \citenamefont
  {Hess}(2008{\natexlab{b}})}]{Boehringer2008247}%
  \BibitemOpen
  \bibfield  {author} {\bibinfo {author} {\bibfnamefont {Klaus}\ \bibnamefont
  {B{\"o}hringer}}\ and\ \bibinfo {author} {\bibfnamefont {Ortwin}\
  \bibnamefont {Hess}},\ }\bibfield  {title} {\enquote {\bibinfo {title} {A
  full time-domain approach to spatio-temporal dynamics of semiconductor
  lasers. ii. spatio-temporal dynamics},}\ }\href {\doibase
  http://dx.doi.org/10.1016/j.pquantelec.2008.10.003} {\bibfield  {journal}
  {\bibinfo  {journal} {Progress in Quantum Electronics}\ }\textbf {\bibinfo
  {volume} {32}},\ \bibinfo {pages} {247--307} (\bibinfo {year}
  {2008}{\natexlab{b}})}\BibitemShut {NoStop}%
\bibitem [{\citenamefont {Wille}\ \emph {et~al.}(2016)\citenamefont {Wille},
  \citenamefont {Sturm}, \citenamefont {Michalsky}, \citenamefont {Röder},
  \citenamefont {Ronning}, \citenamefont {Schmidt-Grund},\ and\ \citenamefont
  {Grundmann}}]{0957-4484-27-22-225702}%
  \BibitemOpen
  \bibfield  {author} {\bibinfo {author} {\bibfnamefont {Marcel}\ \bibnamefont
  {Wille}}, \bibinfo {author} {\bibfnamefont {Chris}\ \bibnamefont {Sturm}},
  \bibinfo {author} {\bibfnamefont {Tom}\ \bibnamefont {Michalsky}}, \bibinfo
  {author} {\bibfnamefont {Robert}\ \bibnamefont {Röder}}, \bibinfo {author}
  {\bibfnamefont {Carsten}\ \bibnamefont {Ronning}}, \bibinfo {author}
  {\bibfnamefont {Rüdiger}\ \bibnamefont {Schmidt-Grund}}, \ and\ \bibinfo
  {author} {\bibfnamefont {Marius}\ \bibnamefont {Grundmann}},\ }\bibfield
  {title} {\enquote {\bibinfo {title} {Carrier density driven lasing dynamics
  in zno nanowires},}\ }\href
  {http://stacks.iop.org/0957-4484/27/i=22/a=225702} {\bibfield  {journal}
  {\bibinfo  {journal} {Nanotechnology}\ }\textbf {\bibinfo {volume} {27}},\
  \bibinfo {pages} {225702} (\bibinfo {year} {2016})}\BibitemShut {NoStop}%
\bibitem [{\citenamefont {Crosignani}\ \emph {et~al.}(1981)\citenamefont
  {Crosignani}, \citenamefont {Porto},\ and\ \citenamefont
  {Papas}}]{Crosignani:81}%
  \BibitemOpen
  \bibfield  {author} {\bibinfo {author} {\bibfnamefont {B.}~\bibnamefont
  {Crosignani}}, \bibinfo {author} {\bibfnamefont {P.~Di}\ \bibnamefont
  {Porto}}, \ and\ \bibinfo {author} {\bibfnamefont {C.~H.}\ \bibnamefont
  {Papas}},\ }\bibfield  {title} {\enquote {\bibinfo {title} {Coupled-mode
  theory approach to nonlinear pulse propagation in optical fibers},}\ }\href
  {\doibase 10.1364/OL.6.000061} {\bibfield  {journal} {\bibinfo  {journal}
  {Opt. Lett.}\ }\textbf {\bibinfo {volume} {6}},\ \bibinfo {pages} {61--63}
  (\bibinfo {year} {1981})}\BibitemShut {NoStop}%
\bibitem [{\citenamefont {Snyder}\ and\ \citenamefont
  {Love}(1983)}]{snyder1983optical}%
  \BibitemOpen
  \bibfield  {author} {\bibinfo {author} {\bibfnamefont {A.W.}\ \bibnamefont
  {Snyder}}\ and\ \bibinfo {author} {\bibfnamefont {J.}~\bibnamefont {Love}},\
  }\href {http://books.google.co.in/books?id=gIQB\_hzB0SMC} {\emph {\bibinfo
  {title} {Optical Waveguide Theory}}},\ Science paperbacks\ (\bibinfo
  {publisher} {Springer},\ \bibinfo {year} {1983})\BibitemShut {NoStop}%
\bibitem [{\citenamefont {H\"{u}gel}\ \emph {et~al.}(2000)\citenamefont
  {H\"{u}gel}, \citenamefont {Heinrich},\ and\ \citenamefont
  {Wegener}}]{Hugel2000}%
  \BibitemOpen
  \bibfield  {author} {\bibinfo {author} {\bibfnamefont {WA}~\bibnamefont
  {H\"{u}gel}}, \bibinfo {author} {\bibfnamefont {MF}~\bibnamefont {Heinrich}},
  \ and\ \bibinfo {author} {\bibfnamefont {M}~\bibnamefont {Wegener}},\
  }\bibfield  {title} {\enquote {\bibinfo {title} {Dephasing due to
  carrier-carrier scattering in 2d},}\ }\href
  {http://onlinelibrary.wiley.com/doi/10.1002/1521-3951(200009)221:1\%3C473::AID-PSSB473\%3E3.0.CO;2-I/abstract}
  {\bibfield  {journal} {\bibinfo  {journal} {physica status solidi (b)}\
  }\textbf {\bibinfo {volume} {473}},\ \bibinfo {pages} {473--476} (\bibinfo
  {year} {2000})}\BibitemShut {NoStop}%
\bibitem [{\citenamefont {Haug}(2000)}]{PSSB:PSSB179}%
  \BibitemOpen
  \bibfield  {author} {\bibinfo {author} {\bibfnamefont {H.}~\bibnamefont
  {Haug}},\ }\bibfield  {title} {\enquote {\bibinfo {title} {Coulomb quantum
  kinetics for semiconductor femtosecond spectroscopy},}\ }\href {\doibase
  10.1002/1521-3951(200009)221:1<179::AID-PSSB179>3.0.CO;2-6} {\bibfield
  {journal} {\bibinfo  {journal} {physica status solidi (b)}\ }\textbf
  {\bibinfo {volume} {221}},\ \bibinfo {pages} {179--188} (\bibinfo {year}
  {2000})}\BibitemShut {NoStop}%
\bibitem [{\citenamefont {Andreasen}\ and\ \citenamefont
  {Cao}(2009)}]{Andreasen:09}%
  \BibitemOpen
  \bibfield  {author} {\bibinfo {author} {\bibfnamefont {Jonathan}\
  \bibnamefont {Andreasen}}\ and\ \bibinfo {author} {\bibfnamefont {Hui}\
  \bibnamefont {Cao}},\ }\bibfield  {title} {\enquote {\bibinfo {title}
  {Finite-difference time-domain formulation of stochastic noise in macroscopic
  atomic systems},}\ }\href
  {http://jlt.osa.org/abstract.cfm?URI=jlt-27-20-4530} {\bibfield  {journal}
  {\bibinfo  {journal} {J. Lightwave Technol.}\ }\textbf {\bibinfo {volume}
  {27}},\ \bibinfo {pages} {4530--4535} (\bibinfo {year} {2009})}\BibitemShut
  {NoStop}%
\bibitem [{\citenamefont {Andreasen}\ and\ \citenamefont
  {Cao}(2010)}]{PhysRevA.82.063835}%
  \BibitemOpen
  \bibfield  {author} {\bibinfo {author} {\bibfnamefont {Jonathan}\
  \bibnamefont {Andreasen}}\ and\ \bibinfo {author} {\bibfnamefont {Hui}\
  \bibnamefont {Cao}},\ }\bibfield  {title} {\enquote {\bibinfo {title}
  {Numerical study of amplified spontaneous emission and lasing in random
  media},}\ }\href {\doibase 10.1103/PhysRevA.82.063835} {\bibfield  {journal}
  {\bibinfo  {journal} {Phys. Rev. A}\ }\textbf {\bibinfo {volume} {82}},\
  \bibinfo {pages} {063835} (\bibinfo {year} {2010})}\BibitemShut {NoStop}%
\bibitem [{\citenamefont {Huang}\ and\ \citenamefont {Ho}(2006)}]{Huang:06}%
  \BibitemOpen
  \bibfield  {author} {\bibinfo {author} {\bibfnamefont {Yingyan}\ \bibnamefont
  {Huang}}\ and\ \bibinfo {author} {\bibfnamefont {Seng-Tiong}\ \bibnamefont
  {Ho}},\ }\bibfield  {title} {\enquote {\bibinfo {title} {Computational model
  of solid-state, molecular, or atomic media for fdtd simulation based on a
  multi-level multi-electron system governed by pauli exclusion and fermi-dirac
  thermalization with application to semiconductor photonics},}\ }\href
  {\doibase 10.1364/OE.14.003569} {\bibfield  {journal} {\bibinfo  {journal}
  {Opt. Express}\ }\textbf {\bibinfo {volume} {14}},\ \bibinfo {pages}
  {3569--3587} (\bibinfo {year} {2006})}\BibitemShut {NoStop}%
\bibitem [{\citenamefont {Thomas}\ and\ \citenamefont
  {Hopfield}(1959)}]{PhysRev.116.573}%
  \BibitemOpen
  \bibfield  {author} {\bibinfo {author} {\bibfnamefont {D.~G.}\ \bibnamefont
  {Thomas}}\ and\ \bibinfo {author} {\bibfnamefont {J.~J.}\ \bibnamefont
  {Hopfield}},\ }\bibfield  {title} {\enquote {\bibinfo {title} {Exciton
  spectrum of cadmium sulfide},}\ }\href {\doibase 10.1103/PhysRev.116.573}
  {\bibfield  {journal} {\bibinfo  {journal} {Phys. Rev.}\ }\textbf {\bibinfo
  {volume} {116}},\ \bibinfo {pages} {573--582} (\bibinfo {year}
  {1959})}\BibitemShut {NoStop}%
\bibitem [{\citenamefont {Thomas}\ and\ \citenamefont
  {Hopfield}(1962)}]{PhysRev.128.2135}%
  \BibitemOpen
  \bibfield  {author} {\bibinfo {author} {\bibfnamefont {D.~G.}\ \bibnamefont
  {Thomas}}\ and\ \bibinfo {author} {\bibfnamefont {J.~J.}\ \bibnamefont
  {Hopfield}},\ }\bibfield  {title} {\enquote {\bibinfo {title} {Optical
  properties of bound exciton complexes in cadmium sulfide},}\ }\href {\doibase
  10.1103/PhysRev.128.2135} {\bibfield  {journal} {\bibinfo  {journal} {Phys.
  Rev.}\ }\textbf {\bibinfo {volume} {128}},\ \bibinfo {pages} {2135--2148}
  (\bibinfo {year} {1962})}\BibitemShut {NoStop}%
\bibitem [{\citenamefont {Maslov}\ and\ \citenamefont
  {Ning}(2003)}]{:/content/aip/journal/apl/83/6/10.1063/1.1599037}%
  \BibitemOpen
  \bibfield  {author} {\bibinfo {author} {\bibfnamefont {A.~V.}\ \bibnamefont
  {Maslov}}\ and\ \bibinfo {author} {\bibfnamefont {C.~Z.}\ \bibnamefont
  {Ning}},\ }\bibfield  {title} {\enquote {\bibinfo {title} {Reflection of
  guided modes in a semiconductor nanowire laser},}\ }\href {\doibase
  http://dx.doi.org/10.1063/1.1599037} {\bibfield  {journal} {\bibinfo
  {journal} {Applied Physics Letters}\ }\textbf {\bibinfo {volume} {83}},\
  \bibinfo {pages} {1237--1239} (\bibinfo {year} {2003})}\BibitemShut {NoStop}%
\bibitem [{\citenamefont {Röder}\ \emph {et~al.}(2015)\citenamefont {Röder},
  \citenamefont {Sidiropoulos}, \citenamefont {Tessarek}, \citenamefont
  {Christiansen}, \citenamefont {Oulton},\ and\ \citenamefont
  {Ronning}}]{doi:10.1021/acs.nanolett.5b01271}%
  \BibitemOpen
  \bibfield  {author} {\bibinfo {author} {\bibfnamefont {Robert}\ \bibnamefont
  {Röder}}, \bibinfo {author} {\bibfnamefont {Themistoklis P.~H.}\ \bibnamefont
  {Sidiropoulos}}, \bibinfo {author} {\bibfnamefont {Christian}\ \bibnamefont
  {Tessarek}}, \bibinfo {author} {\bibfnamefont {Silke}\ \bibnamefont
  {Christiansen}}, \bibinfo {author} {\bibfnamefont {Rupert~F.}\ \bibnamefont
  {Oulton}}, \ and\ \bibinfo {author} {\bibfnamefont {Carsten}\ \bibnamefont
  {Ronning}},\ }\bibfield  {title} {\enquote {\bibinfo {title} {Ultrafast
  dynamics of lasing semiconductor nanowires},}\ }\href {\doibase
  10.1021/acs.nanolett.5b01271} {\bibfield  {journal} {\bibinfo  {journal}
  {Nano Letters}\ }\textbf {\bibinfo {volume} {15}},\ \bibinfo {pages}
  {4637--4643} (\bibinfo {year} {2015})},\ \bibinfo {note} {pMID: 26086355},\
  \Eprint {http://arxiv.org/abs/http://dx.doi.org/10.1021/acs.nanolett.5b01271}
  {http://dx.doi.org/10.1021/acs.nanolett.5b01271} \BibitemShut {NoStop}%
\bibitem [{\citenamefont {R\"{o}der}\ \emph {et~al.}(2016)\citenamefont
  {R\"{o}der}, \citenamefont {Sidiropoulos}, \citenamefont {Buschlinger},
  \citenamefont {Riediger}, \citenamefont {Peschel}, \citenamefont {Oulton},\
  and\ \citenamefont {Ronning}}]{doi:10.1021/acs.nanolett.6b00811}%
  \BibitemOpen
  \bibfield  {author} {\bibinfo {author} {\bibfnamefont {Robert}\ \bibnamefont
  {R\"{o}der}}, \bibinfo {author} {\bibfnamefont {Themistoklis P.~H.}\
  \bibnamefont {Sidiropoulos}}, \bibinfo {author} {\bibfnamefont {Robert}\
  \bibnamefont {Buschlinger}}, \bibinfo {author} {\bibfnamefont {Max}\
  \bibnamefont {Riediger}}, \bibinfo {author} {\bibfnamefont {Ulf}\
  \bibnamefont {Peschel}}, \bibinfo {author} {\bibfnamefont {Rupert~F.}\
  \bibnamefont {Oulton}}, \ and\ \bibinfo {author} {\bibfnamefont {Carsten}\
  \bibnamefont {Ronning}},\ }\bibfield  {title} {\enquote {\bibinfo {title}
  {Mode switching and filtering in nanowire lasers},}\ }\href {\doibase
  10.1021/acs.nanolett.6b00811} {\bibfield  {journal} {\bibinfo  {journal}
  {Nano Letters}\ }\textbf {\bibinfo {volume} {16}},\ \bibinfo {pages}
  {2878--2884} (\bibinfo {year} {2016})}\BibitemShut {NoStop}%
\end{thebibliography}

%

\end{document}